\newcommand{\bE}{\mbox{\boldmath $E$}}

\documentclass[12pt,preprint]{aastex}

\shorttitle{Eight-Octant Phase-Mask Coronagraph}
\shortauthors{N. Murakami et al.}

\begin{document}

\title{An Eight-Octant Phase-Mask Coronagraph}
\author{Naoshi Murakami} 
\affil{National Astronomical Observatory of Japan, 2-21-1 Osawa, Mitaka, Tokyo 181-8588, Japan; }
\email{naoshi.murakami@nao.ac.jp} 

\author{Ryosuke Uemura, Naoshi Baba} 
\affil{Division of Applied Physics, Hokkaido University, Sapporo 060-8628, Japan} 
 
\author{Jun Nishikawa, Motohide Tamura} 
\affil{National Astronomical Observatory of Japan, 2-21-1 Osawa, Mitaka, Tokyo 181-8588, Japan} 

\author{Nobuyuki Hashimoto} 
\affil{Citizen Technology Center Co., 840 Shimotomi, Tokorozawa, Saitama 359-8511, Japan} 
\and 
\author{Lyu Abe}  
\affil{Laboratoire Hippolyte Fizeau, UMR 6525, Universit$\acute {\rm e}$ de Nice-Sophia Antipolis, 28, avenue Valrose, F-06108 Nice, France}

\begin{abstract} 
We present numerical simulations and laboratory experiments on an eight-octant phase-mask (EOPM) coronagraph. 
The numerical simulations suggest that an achievable contrast for the EOPM coronagraph can be greatly improved 
as compared to that of a four-quadrant phase-mask (FQPM) coronagraph for a partially resolved star. 
On-sky transmission maps reveal that the EOPM coronagraph has relatively high optical throughput, a small inner working angle and large discovery space. 
We have manufactured an eight-segment phase mask utilizing a nematic liquid-crystal device, 
which can be easily switched between the FQPM and the EOPM modes. 
The laboratory experiments demonstrate that the EOPM coronagraph has a better tolerance of the tip-tilt error than the FQPM one. 
We also discuss feasibility of a fully achromatic and high-throughput EOPM coronagraph 
utilizing a polarization interferometric technique.
\end{abstract}


\section{Introduction}									
An attempt to discover and characterize extrasolar Earth-like planets is one of the most challenging issues in modern astronomy. 
For direct detection of the extrasolar planets, many concepts of high-contrast imagers have been proposed, 
especially in the context of the Terrestrial Planet Finder Coronagraph (TPF-C) mission. 
It is necessary for the high-contrast imagers to realize extremely high dynamic range over a broad spectral bandwidth, 
together with high optical throughput, a small inner working angle (IWA), and large discovery space. 
For a practical aspect, in addition, technically simple instrumentation would be preferable because of a severe 
environment of space observations.

A classical Lyot coronagraph is a precursory work for the high-contrast imager aiming at observing a solar coronal prominence 
by using an opaque mask on a focal plane \citep{Lyot39}. 
Recently, more advanced focal-plane masks, such as a four-quadrant phase mask (FQPM), a band-limited mask, and 
an optical vortex mask, have been proposed for planet detection \citep{Rouan00,Kuchner02,Kuchner05,Foo05}. 
These techniques have the ability to strongly suppress both stellar diffraction core and halo, and tend to have relatively small IWAs. 
Pupil-apodization techniques, such as an apodized square aperture and binary shaped masks, 
have been proposed to suppress only the stellar diffraction halo \citep{Nisenson01,Kasdin03,Enya07}. 
An apodized pupil Lyot coronagraph (APLC), a combination of the pupil-apodization and the classical Lyot coronagraph, 
has also been proposed for realizing the high-contrast imaging with arbitrary aperture shapes of telescopes \citep{Soummer05}. 
The pupil-apodizations are less sensitive to tip-tilt errors and/or finite stellar angular sizes than the focal-plane mask coronagraphs. 
However, these techniques tend to have relatively large IWAs and low optical throughputs. 
A phase-induced amplitude apodization (PIAA) has been proposed to achieve a small IWA without a loss of 
the optical throughput \citep{Guyon03}.

A comprehensive evaluation of theoretical capabilities for extrasolar planet detection 
has been conducted for several promising concepts \citep{Guyon06}. 
For direct detection of the extrasolar Earth-like planets, the high-contrast imagers must be 
tolerant of the tip-tilt error and the stellar angular size problem. 
For the FQPM coronagraph, however, residual stellar noise increases rapidly with $\delta^2$, where $\delta$ is the tip-tilt error. 
This second-order behavior for the tip-tilt error would be insufficient for direct detection of the Earth-like planets around nearby stars. 
For this reason, the FQPM coronagraph has been excluded from the candidates for the TPF-C mission. 
Note that it is possible to realize the higher-order insensitivities to the tip-tilt error by using the other focal-plane mask 
(the band-limited mask and the optical vortex mask) coronagraphs. 
Nevertheless, the FQPM coronagraph is still attractive because of its ability to realize a perfect stellar suppression, 
a high optical throughput, a large discovery space and a small IWA together with technically simple instrumentation. 
\cite{Rouan07} point out that it is possible to realize more efficient stellar suppressions by multiplying the number of sectors of the phase mask. 
We have also been paying attention to an eight-octant phase mask (EOPM) because of its acceptable optical throughput, IWA, and discovery space. 
We expect that it is easy to manufacture and achromatize the high-performance EOPM, because it has neither continuous phase retardation 
like the optical vortex masks nor complex and fine structure like the band-limited masks.

We have manufactured a chromatic version of the phase mask by using a nematic liquid crystal device, which can be easily switched between the FQPM and the EOPM modes. 
In this paper, we present numerical simulations and preliminary laboratory experiments on the EOPM coronagraph comparing it with the FQPM one. 
In \S. \ref{sec:EOPM}, we show the results of the coronagraphic numerical simulations for a partially resolved star. 
On-sky transmittances of both the coronagraphs are also shown for evaluating their optical throughputs, IWAs, and discovery spaces. 
In \S. \ref{sec:Lab}, we describe the manufactured phase mask, and show results of the laboratory experiments 
on both the coronagraphs for various tip-tilt errors. 
In \S. \ref{sec:Achromatization}, we discuss feasibility of a fully achromatic and high-throughput EOPM coronagraph 
utilizing a polarization interferometric technique. 
Finally, our conclusions are summarized in \S. \ref{sec:Conc}.

\section{Numerical Simulations}							
\label{sec:EOPM}
Similar to the FQPM, the EOPM is put on a focal plane to divide a stellar image into eight-octant regions, 
and provides a $\pi$-phase difference between the adjacent octants. 
This causes a self-destructive interference inside the pupil area on a following reimaged pupil plane, 
where a Lyot stop is put to block stellar light diffracted outside the pupil. 
A transmittance of the EOPM is defined as 
\begin{equation}
M(\psi) = (-1)^k \qquad \frac{k \pi}{4} \le \psi < \frac{(k+1)\pi}{4}, 
\end{equation}
where $\psi$ is an azimuth angle in the focal plane, and $k$ is an arbitrary integer. 
The transmittance can be expressed by using a Fourier expansion as 
\begin{equation}
M(\psi) = \frac{2}{\pi i} \left( \sum_{m=0}^{\infty} \frac{e^{4(2m+1)i\psi}}{2m+1} - 
							\sum_{m=0}^{\infty} \frac{e^{-4(2m+1)i\psi}}{2m+1} \right). 
\end{equation}
The above equation shows that the EOPM can be written as a weighted sum of evenly charged optical vortex masks, 
each of which causes zero intensity within the pupil in a Lyot-stop plane for a pointlike star \citep{Jenkins08}. 
Thus, the EOPM coronagraph realizes perfect stellar elimination for unresolved stars as well as the FQPM one does. 
In practice, however, observed stars will have finite angular sizes, and residual stellar noise will increase due to off-axis light 
from a rim of the stellar disk.

\begin{figure}
\epsscale{1.0}
\plotone{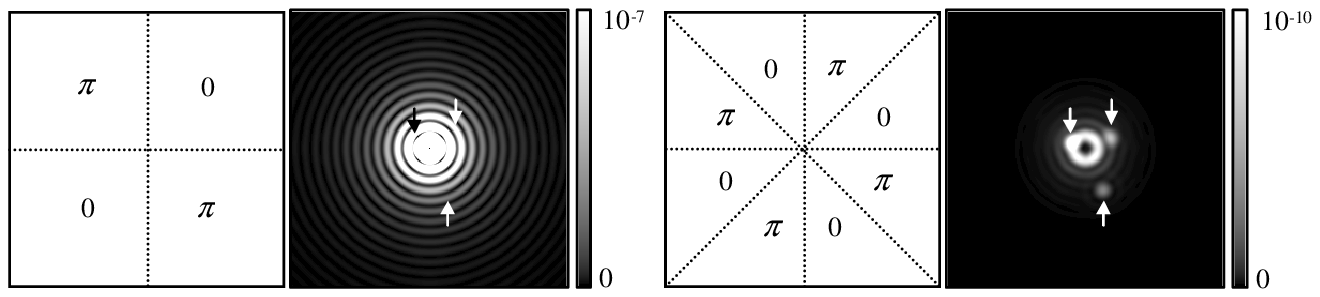}
\caption{
Drawings of phase-masks and simulated coronagraphic images for FQPM ({\it left}) and EOPM coronagraphs ({\it right}). 
A partially resolved central star (angular diameter of 0.02 $\lambda/D$) and 
three model planets (indicated by arrows) are assumed in the numerical simulation. 
}\label{fig:SimImg}
\end{figure}

We carry out numerical simulations for evaluating coronagraphic performance of both the coronagraphs for partially resolved stars. 
In Figure \ref{fig:SimImg}, we show the results of the numerical simulations, together with illustrations of the FQPM and the EOPM. 
In the numerical simulations, we assume a uniform-disk star with an angular diameter of 0.02 $\lambda/D$ as an example. 
This angular size roughly corresponds to a Sun-like star seen from 
10 parsec away from the Earth assuming $D = $3 m and $\lambda = $ 0.6 $\mu$m. 
A Lyot-stop size is set to be 90\% of an entrance pupil. 
Two images show the coronagraphic images for a model planetary system around the partially resolved star. 
The model planetary system is composed of three model planets at 1.5, 3.0, and 5.0 $\lambda/D$ from the central star with 
star/planet intensity ratios of $2\times10^{9}$, $1\times10^{10}$, and $2\times10^{10}$, respectively. 
The model planetary images can be clearly detected by the EOPM coronagraph as pointed by arrows, 
while these images are buried in the residual stellar noise for the FQPM-coronagraphic image. 
Note that instrumental defects (chromatism, phase and amplitude aberrations, misalignment, a pointing error of a telescope, 
an effect of a central obscuration of a secondary mirror, and so on) are not included in the numerical simulations.

\begin{figure}
\epsscale{1.0}
\plotone{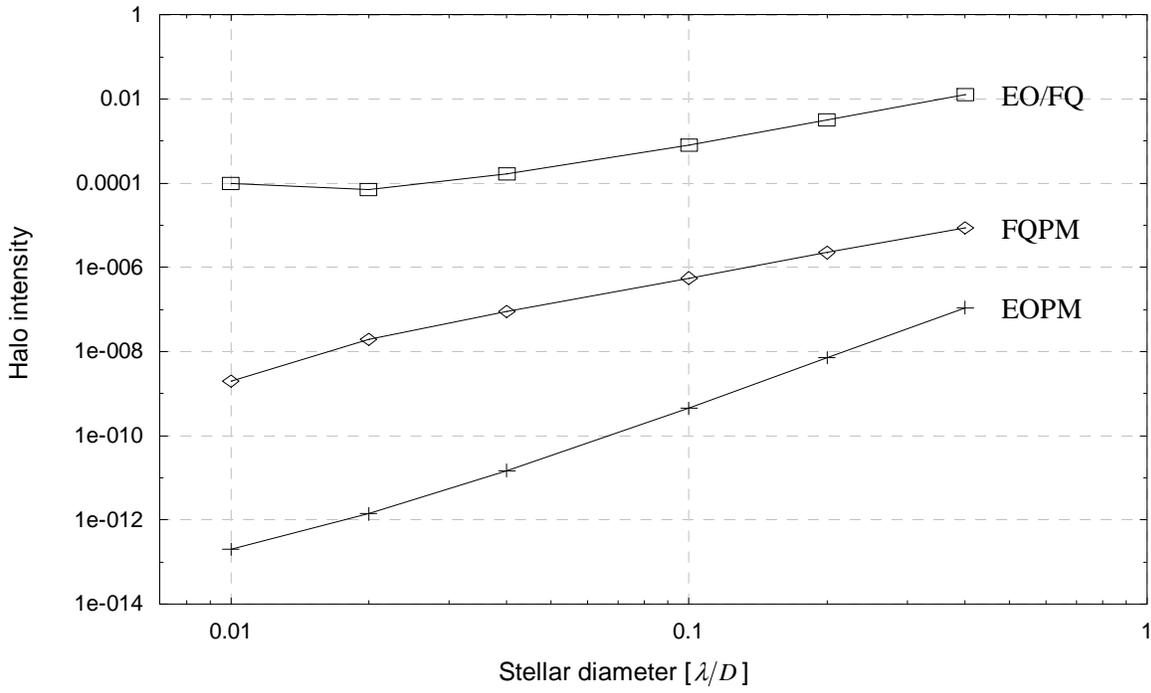}
\caption{
Normalized halo intensity (mean intensity over 4-6 $\lambda/D$) as 
a function of a stellar diameter in unit of $\lambda/D$. 
Ratios between FQPM and EOPM are also shown. 
}
\label{fig:HalovsDia}
\end{figure}

We also evaluate coronagraphic performance for partially resolved stars of various angular diameters. 
Figure \ref{fig:HalovsDia} shows halo intensity (mean intensity over 4-6 $\lambda/D$) 
as a function of the stellar diameter in $\lambda/D$ unit. 
A ratio of the halo intensities of both the coronagraphs is also shown in the graph. 
The results suggest that the coronagraphic performance can be greatly improved by using the EOPM especially for smaller stars. 
For example, the coronagraphic performance can be $10^4$ times improved for stars with a diameter less than 0.04 $\lambda/D$, 
and a contrast greater than $10^9$ will be achieved at the angular distance around 5 $\lambda/D$ even for a stellar diameter of 0.1 $\lambda/D$.

\begin{figure}
\epsscale{1.0}
\plotone{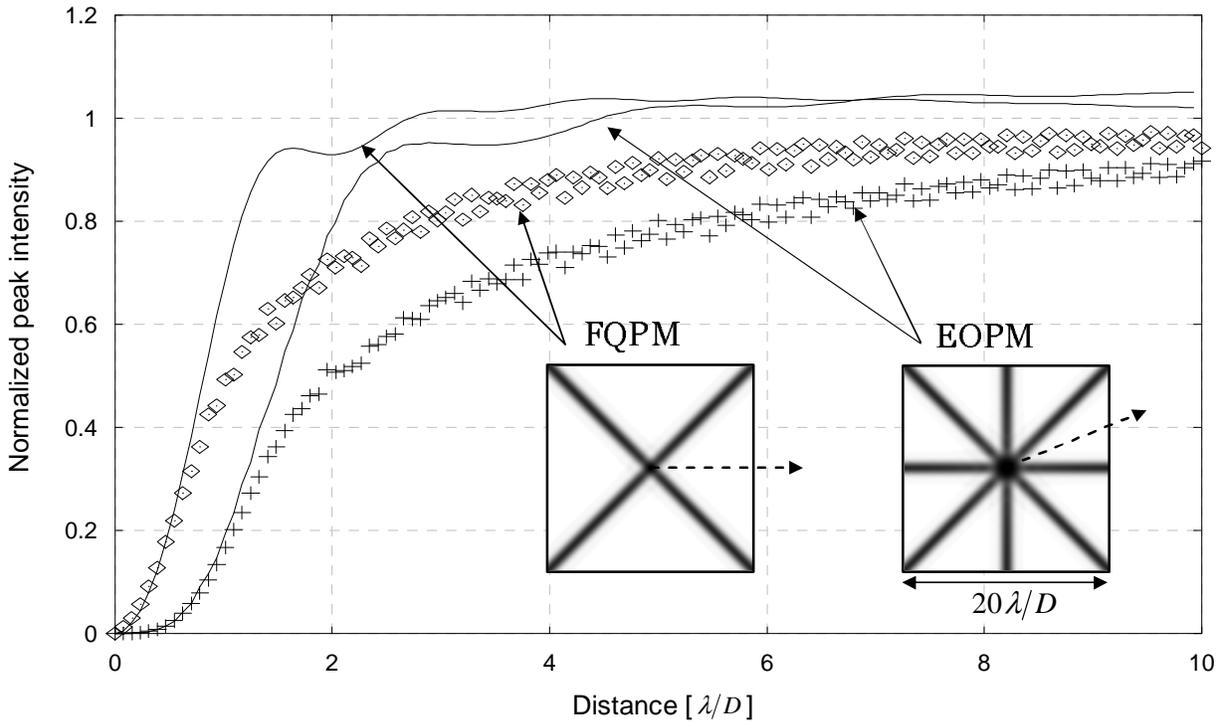}
\caption{Calculated on-sky transmission maps for the FQPM and EOPM coronagraphs (a field-of-view of $20\times20$ $\lambda/D$). 
Two plots in the graph are radial profiles of the transmission maps (diamonds for the FQPM, and plus signs for the EOPM), 
while two curves show the profiles along the high-throughput axes (shown by dashed arrows in the maps).
}
\label{fig:TMap}
\end{figure}

Next, we show on-sky transmission maps for both the coronagraphs to evaluate their IWAs and discovery spaces. 
Figure \ref{fig:TMap} shows the transmission maps with a field of view of $20\times20$ $\lambda/D$, 
in which we calculate peak intensities of an off-axis light source from each on-sky position normalized by those without the phase masks. 
Two plots (shown by diamonds and crosses) in Figure \ref{fig:TMap} show the azimuthally averaged transmission 
for both the coronagraphs as a function of an angular distance in unit of $\lambda/D$ from a center of the phase mask. 
Two curves in the graph show the transmittances along "high-throughput axes" between the boundaries of the phase masks. 
The high-throughput axes for both the phase masks are shown by dashed arrows in each transmission map. 
Note that the peak intensities exceed unity because a loss of intensity due to a Lyot stop is not considered in these results. 
The loss due to the Lyot stop becomes $D_L^4$, where $D_L$ is a relative size of the Lyot stop to an entrance pupil. 
In the case of the numerical simulation shown here, the loss becomes $D_L^4=0.66$ because we set the Lyot-stop size to $D_L=0.9$.

From the transmission maps, we conclude that the FQPM has an IWA of 1.1 $\lambda/D$ while that of the EOPM is 1.9 $\lambda/D$. 
Here, we define the IWA as an angular distance at which the transmission along the high-throughput axis including the loss due to 
the Lyot stop reaches 50\%. 
Note that the transmission of 50\% corresponds to 0.76 in Figure \ref{fig:TMap}, 
because the numerical simulations do not include the loss due to the Lyot stop. 
Although we estimate the IWA of the EOPM as 1.9 $\lambda/D$ here, 
faint companions imaged at inner region (e.g., 1.5 $\lambda/D$) would be detectable as demonstrated in Figure \ref{fig:SimImg}.

We also evaluate the discovery spaces as 81\% and 71\% for the FQPM and EOPM coronagraphs, respectively. 
Here, we define the discovery space as a fractional region, with respect to a full field of view of $20\times20$ $\lambda/D$, 
in which the transmission exceeds 50\%. 
Note that planetary images are strongly suppressed on the boundary of the phase masks, and thus rotations of the phase masks 
($45^\circ$ for the FQPM and $22.5^\circ$ for the EOPM) are required for obtaining a full-sky coverage.

\section{Laboratory Demonstrations}						
\label{sec:Lab}
\begin{figure}
\epsscale{1.0}
\plotone{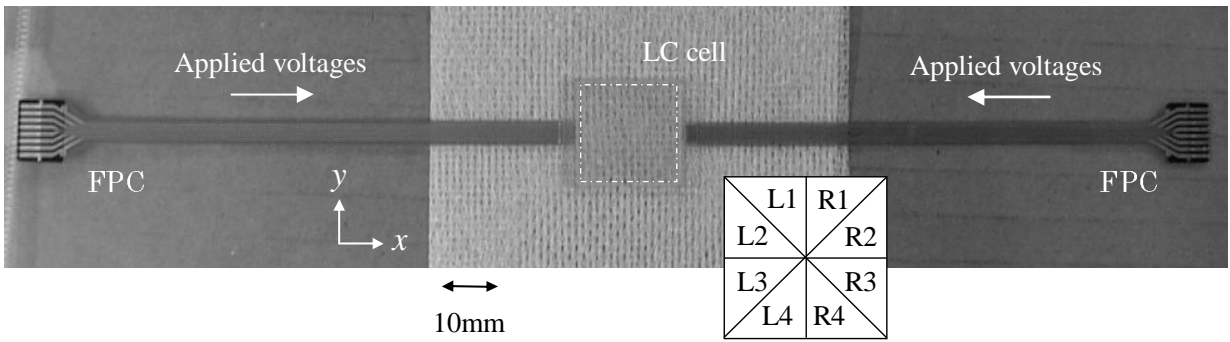}
\caption{A picture of a manufactured NLC phase mask. 
A liquid-crystal cell is subdivided into eight segments ($R1$-$R4$ and $L1$-$L4$, as shown by an inset), to which 
the flexible printed circuits (FPCs) are connected for applying voltages separately. 
A coronagraphic phase mask can be realized for a linearly polarized light along $x$-direction 
by adjusting the applied voltages appropriately.}
\label{fig:LCMask}
\end{figure}

We have manufactured an eight-segment phase mask utilizing a nematic liquid crystal (NLC) device with homogeneous alignment. 
Figure \ref{fig:LCMask} is a picture of the manufactured phase mask. 
The NLC device is subdivided into eight segments, $L1$-$L4$ and $R1$-$R4$, 
which can be connected to function generators separately via flexible printed circuits (FPCs) 
for applying desired voltages to each segment. 
When a linearly polarized (LP) light along $x$-direction (parallel to the alignment of the NLC) enters, 
the resultant polarization is unchanged, while the phase can be modulated by the applied voltage $V$. 
The phase of the resultant LP light can be written as $\phi(V) = 2 \pi n(V) d/\lambda$, where $n$ and $d$ are 
a refractive index and thickness of the NLC device, and $\lambda$ is an operational wavelength. 
For realizing a coronagraphic phase-mask, the applied voltages $V_a$ and $V_b$ to the appropriate segments 
(e.g., $V_a$ to segments [$L1, L3, R2, R4$] and $V_b$ to [$L2, L4, R1, R3$] for the EOPM-mode) must be adjusted 
so that the phase difference $\Delta \phi = \phi(V_a)-\phi(V_b)$ is $(2m+1)\pi$ ($m$ is an arbitrary integer).

In our laboratory demonstrations, we choose the applied voltages (a rectangular wave of 1 kHz) of 
$V_b=2.73$ V rms and $V_a=0.0$ V rms for realizing $\Delta \phi=3\pi$ (i.e. $m=1$). 
It should be noted that the manufactured phase mask cannot be used for a broadband light because of a chromatic 
property of the NLC device although it can be optimized for an arbitrary wavelength by adjusting the applied voltages. 
An issue of an achromatization of the phase-mask will be discussed in the following section.

\begin{figure}
\epsscale{1.0}
\plotone{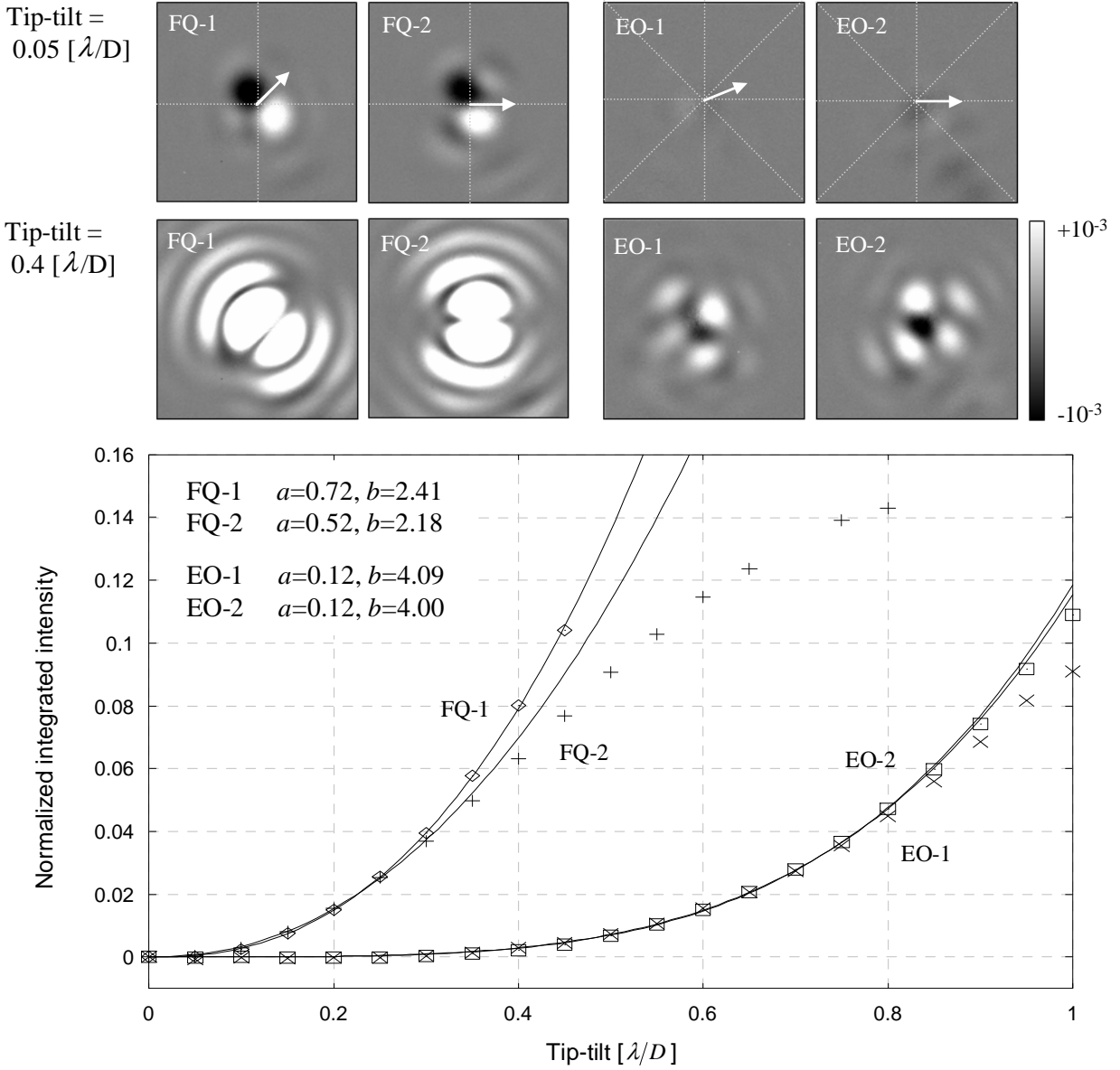}
\caption{
Top figures show the acquired FQPM- and EOPM-coronagraphic images with tip-tilt errors of 0.05 ({\it top}) and 0.4 $\lambda/D$ ({\it bottom}). 
The tip-tilt directions and boundaries of the phase-masks are shown by white arrows and dotted lines in the top images. 
Zero-tip-tilt images are subtracted from each image for suppressing residual speckle noise. 
Plots in a bottom graph show normalized intensities $I(\delta)$ integrated over a field of view as a function of the tip-tilt error $\delta$. 
Curves show fitting results of each data with a function $I(\delta)=a\delta^b$. 
Fitted parameters $a,b$ for each situation are also shown in the graph.} 
\label{fig:TTPerf}
\end{figure}

We have carried out laboratory demonstrations of the FQPM and the EOPM coronagraphs with a monochromatic light source 
(He-Ne laser with $\lambda=0.633$ $\mu$m) 
to acquire coronagraphic images with various tip-tilt errors. 
Figure \ref{fig:TTPerf} shows experimental results. 
The top panels are acquired images with the tip-tilt errors $\delta$ of 0.05 and 0.4 $\lambda/D$. 
White arrows and dotted lines indicate the tip-tilt directions and the boundaries of the phase masks, respectively. 
We show the experimental results for two tip-tilt directions; one is that between the boundaries (FQ-1 and EO-1) and the other is 
that along the boundary (FQ-2 and EO-2). 
For evaluating a residual intensity due to the tip-tilt error, 
we subtract a zero-tip-tilt coronagraphic image from each image to eliminate residual speckle noise. 
We assume that the residual speckle noise comes especially from an optical surface roughness of the NLC phase mask itself, 
which is measured to be around 0.01 $\lambda$ rms over a diameter of 5.4 mm. 
For on-sky observations, the speckle subtraction will be carried out by subtracting a reference pointlike star 
of a similar spectral type to a target. 
Note that the speckle noise can also be suppressed by using the differential techniques \citep{Racine99,Baba03,Murakami07} or 
the speckle nulling technique \citep{Trauger07}.

We calculate a total intensity $I(\delta)$ of the coronagraphic images over a field of view of 7.4$\times$7.4 $\lambda/D$. 
The bottom graph in Figure \ref{fig:TTPerf} shows the result. 
Four plots are the measured intensities $I(\delta)$ for the FQPM (diamonds and plus signs) and the EOPM (squares and crosses). 
The residual intensity for the FQPM coronagraph rapidly increases with the tip-tilt error, 
while that for the EOPM one can be well suppressed even for large tip-tilt errors. 
We also fit the experimental data with a function $I(\delta)=a\delta^b$, where $a$ and $b$ are fitting parameters. 
The fittings are carried out by using the data for small tip-tilt errors ($\delta<0.3$ for the FQPM and $\delta<0.7$ for the EOPM). 
The fitting results and the fitted parameters for both the coronagraphs with two different tip-tilt directions are 
also shown in Figure \ref{fig:TTPerf}.
Consequently, we obtain the values $b = $2.41 and 2.18 for the FQPM coronagraph, while $b =$ 4.09 and 4.00 for the EOPM one. 
These results demonstrate that the EOPM coronagraph has a tolerance of the tip-tilt error (a fourth-order response) 
as compared to the FQPM one (roughly a second-order response).

\section{Discussion} 		
\label{sec:Achromatization}
The manufactured phase-mask cannot be used for a broadband light because of the chromatic effect of the NLC device. 
We estimate that, when an acceptable phase error is set to $\pm 50$ mrad, 
an effective spectral bandwidth at $\lambda=$650 nm is restricted to only 16 nm (that is, $\Delta \lambda/\lambda=0.025$). 
For our future work, we are planning to design and manufacture an achromatic EOPM. 
We expect that it would be much easier to manufacture the achromatic EOPM than the other focal-plane coronagraphic masks, 
because the EOPM has neither a continuous phase shift nor complex and fine structure.

Many techniques for the achromatic phase mask have been proposed in the context of the FQPM coronagraph 
that can be directly applied to the EOPM one. 
\cite{Riaud01} propose to use a reflective phase mask with a quarter-wave multilayer coating of high- and low-optical indices. 
\cite{Abe01} propose a phase-knife coronagraph, in which a white-light Airy image is dispersed to provide $\pi$-phase shift separately 
for each wavelength. 
A zero-order grating has also been proposed, which makes use of a form birefringence of subwavelength surface-relief gratings \citep{Mawet05}. 
\cite{Mawet06} propose to use a two stage stack of birefringence half-wave plates (HWPs) with two materials. 
\cite{Baudoz08} propose a multistage FQPM coronagraph by constructing chromatic phase-mask coronagraphs in cascade. 
Note that the manufactured NLC-mask might be advantageous to this technique because of its adjustable operational wavelength. 
An application of the manufactured NLC-mask to the multistage FQPM coronagraph will be an interesting future work.

A FQPM by utilizing a polarization interferometry (a four-quadrant polarization-mask; FQPoM) has been proposed \citep{Baba02}, 
and very high contrast has been demonstrated \citep{Murakami08}. 
The FQPoM is composed of four HWPs made of a ferroelectric liquid crystal (FLC) device with different angles of optical axes between quadrants. 
These HWPs rotate an input LP light to opposite directions by $\pm 45^\circ$ \citep{Murakami03}. 
This method requires two polarizers, one in front of and one behind the FQPoM, whose orientation angles are $45^\circ$ and $-45^\circ$, respectively. 
By utilizing the polarization interferometry, it is possible to realize a fully achromatic phase mask 
in spite of the chromatic characteristic of the FLC device. 
However, this original design has very low optical throughput, which is restricted to be at most 0.25 because of the two polarizers 
(intensity losses of 0.5 at each polarizer assuming unpolarized planetary signal).

This problem can be solved by replacing the first $45^\circ$ polarizer with a polarizing beam splitter to construct a two-channel configuration 
for utilizing both $45^\circ$ and $-45^\circ$ LP components (${\rm LP_A}$ and ${\rm LP_B}$, respectively) \citep{Baba03}. 
Furthermore, the optical throughput can also be greatly improved by changing the angles of the LP rotation from $\pm45^\circ$ to $\pm90^\circ$ 
as mentioned in \cite{Murakami06}.

\begin{figure}
\epsscale{1.0}
\plotone{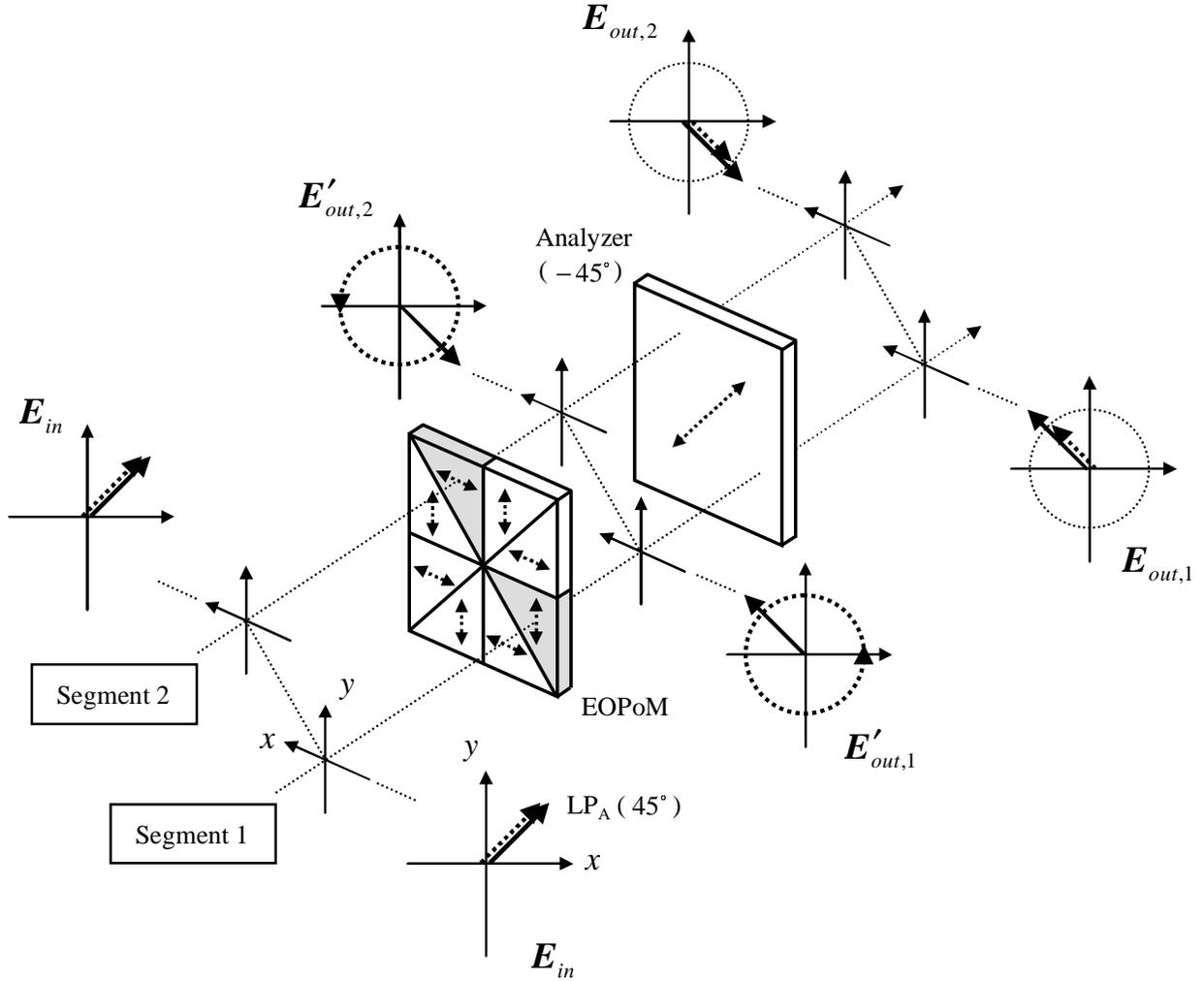}
\caption{A schematic design of a fully achromatic and high-throughput eight-octant polarization mask. 
Implementation of the EOPoM is also shown for input ${\rm LP_A}$ ($45^\circ$ linearly polarized) light beams 
passing through two segments (segment 1 and 2). 
}
\label{fig:EOPoM}
\end{figure}

Figure \ref{fig:EOPoM} shows a fully achromatic and high-throughput design of the EOPM coronagraph, 
which we call an eight-octant polarization mask (EOPoM). 
When the ${\rm LP_A}$ light enters to the phase mask, 
whose Jones vector can be written as $\bE_{in}= \case{1}{\sqrt{2}}[1 \; 1]^\dagger$ ($\dagger$ means transposed matrix), 
Jones vectors of the output beams from two segments of the EOPoM (segments 1 and 2 in Figure \ref{fig:EOPoM}) can be written as 
\begin{equation} 
\bE'_{out,1} = \frac{1}{2\sqrt{2}} 
\left[
\matrix{
1 	\cr
e^{i \beta} 	\cr
}
\right], \;
\bE'_{out,2} = \frac{1}{2\sqrt{2}} 
\left[
\matrix{
e^{i \beta}	\cr
1 	\cr
}
\right]. 
\label{eq:output_pre}
\end{equation} 
Here, $\beta$ is a retardation of the FLC device, which must ideally be $\pi$ (EOPoM is regarded as perfect HWPs), 
but in general, depends strongly on a wavelength. 
Equation (\ref{eq:output_pre}) suggests that both the output light beams become $-45^\circ$ LP light when the retardation $\beta=\pi$. 
On the other hand, the outputs become circular polarizations of opposite directions 
when the EOPoM is regarded as quarter-wave plates (QWPs; $\beta=\pi/2$). 
After passing though the EOPoM, output Jones vectors from the $-45^\circ$ analyzer are calculated as 
\begin{equation} 
\bE_{out,1} = \frac{1}{2\sqrt{2}} 
\left[
\matrix{
1 - e^{i \beta} 	\cr
-1 + e^{i \beta} 	\cr
}
\right], \;
\bE_{out,2} = \frac{1}{2\sqrt{2}} 
\left[
\matrix{
-1 + e^{i \beta}	\cr
1 - e^{i \beta} 	\cr
}
\right]. 
\label{eq:output}
\end{equation} 
Equation (\ref{eq:output}) clearly shows that the resultant light beams satisfy $\bE_{out,2} = e^{i \pi}\bE_{out,1}$ for any values of $\beta$. 
Thus, a fully achromatic phase-mask can be realized despite the chromatic property of the FLC device. 
In Figure \ref{fig:EOPoM}, we describe states of polarization before and after the EOPoM and after the analyzer 
for two cases (solid lines for $\beta=\pi$, and dotted lines for $\beta=\pi/2$), 
assuming the input ${\rm LP_A}$ light beams.

Almost all photons of a planetary image pass through one segment of the phase mask (e.g., segment 1). 
Thus, an optical throughput for the planetary signal can be written as $T=|\bE_{out,1}|^2/|\bE_{in}|^2$, and becomes 
\begin{equation} 
T = \case{1}{2}(1 - \cos \beta). 
\end{equation}
Therefore, a lossless EOPM coronagraph (that is, $T=1.0$) can be realized when $\beta =\pi$, but 
the optical throughput decreases as $\beta$ deviates from $\pi$. 
For achieving high throughput of $T>0.8$, a deviation of the retardation $\beta$ from $\pi$ must be less than $\pm0.92$ radian. 
We measured the retardation $\beta(\lambda)$ of a FLC device used in FQPoM-coronagraphic demonstrations reported in \cite{Murakami08}. 
From the measurements, we roughly estimate that the throughput of $T>0.8$ can be achieved 
for a broad bandwidth of $\Delta \lambda/\lambda=0.4$ in a visible spectral range.

We expect that the achromatic EOPoM can also be realized by utilizing a twisted-nematic liquid crystal (TNLC) device with 
twist angles of $\pm90^\circ$, because this kind of device can also rotate an input LP light to $\pm90^\circ$. 
Evaluations of the achromaticity and the optical throughput of the FLC- and TNLC-based phase-masks will be our future works.

\section{Conclusions} 				
\label{sec:Conc} 
In this paper, we report the numerical simulations and the laboratory experiments on the EOPM coronagraph. 
The numerical simulations show that the EOPM coronagraph greatly improves the achievable contrast as compared to the FQPM 
one for partially resolved stars. 
In addition, on-sky transmission maps reveal that the EOPM coronagraph has a relatively small IWA and a large discovery space. 
We manufactured the eight-segment phase mask by using the NLC device, 
which can be easily switched between the FQPM and EOPM modes. 
Our laboratory experiments confirm that the EOPM coronagraph has the fourth-order response to the tip-tilt error, 
which is better than that of the FQPM one (expected to be second-order). 
This higher-order behavior of the EOPM coronagraph could enable us to detect extrasolar Earth-like planets around nearby stars 
with relatively large apparent sizes.

We also discuss feasibility of a fully achromatic and high-throughput EOPM coronagraph. 
We expect that the polarization interferometric phase mask by using the FLC device (so-called EOPoM) and the two-channel optical configuration to 
improve the achromaticity and the optical throughput. 
The two-channel configuration is also useful for obtaining polarization differential images 
to suppress residual stellar speckle noise and to carry out polarimetric measurements of extrasolar planets \citep{Baba03}. 
For our future work, we are planning to design and manufacture the achromatic EOPoM by using the FLC (or possibly by the TNLC) device, 
and conduct on-sky high-contrast observations of circumstellar disks and faint companions around nearby stars.

\acknowledgments
We thank T. Inabe and H. Shibuya of Hokkaido University for their experimental assistance. 
We are grateful to K. Oka of Hokkaido University for his valuable comments and support. 
We thank the Advanced Technology Center of the National Astronomical Observatory of Japan for helpful support. 
This research was partly supported by the National Astronomical Observatory of Japan 
and by a Grant-in-Aid for Scientific Research (B) from the Japan Society for the Promotion of Science (19360036). 
This research was also partly supported by the Japanese Ministry of Education, Culture, Sports, Science and Technology, 
through a Grant-in-Aid for Scientific Research on Priority Areas, "Development of Extra-solar Planetary Science". 
N. M. is financially supported by the Research Fellowships of the Japan Society for the Promotion of Science for Young Scientists.

\clearpage

\clearpage

\clearpage

\clearpage

\clearpage

\clearpage

\end{document}